\documentclass[aps,twocolumn,showpacs,superscriptaddress,amsmath,amssymb]{revtex4}
\usepackage{graphicx}
\usepackage{dcolumn}
\usepackage{bm}
\begin{document}
\preprint{NUHEP-EXP/07-12}
\title{Enhanced Rare Pion Decays from a Model of MeV Dark Matter}
\author{Yonatan Kahn}
\author{Michael Schmitt}
\affiliation{Northwestern University, Department of Physics and Astronomy,
2145 Sheridan Road, Evanston, IL 60208, USA}
\author{Tim M.P. Tait}
\affiliation{Northwestern University, Department of Physics and Astronomy,
2145 Sheridan Road, Evanston, IL 60208, USA}
\affiliation{HEP Division, Argonne National Lab, Argonne IL 60439 USA}
\newcommand{\keV}                {{\mathrm{keV}}}
\newcommand{\MeV}                {{\mathrm{MeV}}}
\newcommand{\GeV}                {{\mathrm{GeV}}}
\newcommand{\Mchi}               {M_{\chi}}
\newcommand{\Mchisq}             {M_{\chi}^2}
\newcommand{\MU}                 {M_{U}}
\newcommand{\MUsq}               {M_{U}^2}
\newcommand{\Cchi}               {C_\chi}
\newcommand{\Cchisq}             {C_\chi^2}
\newcommand{\fe}                 {f_e}
\newcommand{\fesq}               {f_e^2}
\newcommand{\fnu}                {f_\nu}
\newcommand{\fnusq}              {f_\nu^2}
\newcommand{\Beeann}             {B^{ee}_{\mathrm{ann}}}
\newcommand{\epem}               {e^+e^-}
\newcommand{\mpmm}               {\mu^+\mu^-}
\newcommand{\pion}               {\pi^0}
\newcommand{\decay}              {\pion \to \epem}
\newcommand{\dalitzdecay}        {\pion \to \epem\gamma}
\newcommand{\pb}                 {\mathrm{pb}}

\newcommand{\Gammatot}           {\Gamma_{\mathrm{tot}}}
\newcommand{\BR}                 {B}
\newcommand{\BRmeas}             {B^{\mathrm{meas}}}
\newcommand{\BRSM}               {B^{\mathrm{SM}}}
\newcommand{\bi}   {\begin{itemize}}
\newcommand{\ei}   {\end{itemize}}
\newcommand{\be}   {\begin{enumerate}}
\newcommand{\ee}   {\end{enumerate}}
\newcommand{\bcen} {\begin{center}}
\newcommand{\ecen} {\end{center}}
\newcommand{\beq}  {\begin{equation}}
\newcommand{\eeq}  {\end{equation}}
\newcommand{\bdm}  {\begin{displaymath}}
\newcommand{\edm}  {\end{displaymath}}

\newcommand{\etal} {{\em et al. }}
\newcommand{\ie} {{\em i.e.}}
\newcommand{\eg} {{\em e.g.}}

\date{August 12, 2008}

\begin{abstract}
A model has been proposed in which neutral scalar particles $\chi$, of
mass $1-10~\MeV$, annihilate through the exchange of a light vector
boson $U$, of mass $10-100~\MeV$, to produce the $511~\keV$ line observed
emanating from the center of the galaxy.
The $\chi$ interacts weakly with normal matter and is a viable dark matter candidate.
If the $U$-boson couples to quarks as well as to electrons, it could enhance
the branching ratio for the rare decay $\decay$.  A recent measurement by
the KTeV Collaboration lies three standard deviations above a prediction
by Dorokhov and Ivanov, and we relate this excess to the couplings of
the $U$-boson.  The values are consistent with other constraints and
considerations.  We make some comments on possible improvements in
the data.
\end{abstract}


\maketitle

\section{\label{S:intro}Introduction}
\par
Astronomical observations over the past century have shown that approximately
20\% of the universe is made of dark matter. Since dark matter has only been
detected through its gravitational interactions, several
of its properties, including the mass of dark matter particles and their
interactions with Standard Model (SM) particles, remain completely unknown.
While the WIMP (Weakly Interacting Massive Particle) with mass of order the
electroweak scale is arguably the most popular candidate at
this time, the possibility that dark matter could be lighter than
$100~\MeV$ has attracted much attention recently. A model proposed by
Boehm \etal \cite{BoehmFayet} postulates a neutral scalar dark matter
particle $\chi$ with  mass $1-10~\MeV$ \cite{Beacom}, 
which annihilates to produce
electron/positron pairs:  $\chi \chi \to \epem$
(alternately, $\chi$ could be a fermion  \cite{Rasera:2005sa}).
The excess positrons produced in this annihilation
reaction could be responsible for the bright $511~\keV$ line emanating
from the center of the galaxy~\cite{INTEGRAL}, as more conventional
astrophysical explanations have failed to explain both the intensity and
shape of this line.
Boehm proposes two particles which mediate $\chi$ annihilation: a
neutral vector boson $U$, with mass $m_U \sim 10-100~\MeV$, and a heavy
fermion $F^\pm$ with mass $> 100~\GeV$. The $U$ boson is needed to explain
the relic dark matter density, while the $F$ fermion is necessary to account for
the observed rate of positron annihilation in the galactic center~\cite{Ascasibar}.
For early incarnations of the $U$ boson, see \cite{Fayet:1980rr}.

\par
The rare decay $\decay$ has long posed an interesting problem
in the theory of strong interactions.  Since it is suppressed
at tree-level, the rate is very small, which provides an opening
for indirect effects of new physics to appear.
A new precise measurement of $\BR(\decay)$ by the KTeV Collaboration
exceeds the most recent theoretical calculation.  In this Letter, we
examine the possibility that this excess can be explained by the
exchange of an off-shell $U$ boson of the type proposed in the
light dark matter model, and we compare the constraints obtained
on the coupling constants to other constraints.

\section{\label{S:pion}Pion decay}
\par
The tree-level decay $\pion \to \gamma^* \to \epem$ is forbidden
because the pion is a pseudo-scalar particle and the
massless photon has no longitudinal component.  Consequently,
the lowest-order SM contribution is a one-loop
process with a two-photon intermediate state.
The suppression of the amplitude by a factor of $\alpha^2$
and by helicity conservation
leads to an extremely small decay width, thus allowing
for even tiny effects of new physics to be detectable.
\par
The KTeV-E779 Collaboration recently published a new measurement
of $\BR(\decay)$~\cite{KTEV}.  They normalized this branching
ratio to the Dalitz decay mode $\BR(\dalitzdecay)$, taking both
sets of decays from a large sample of $K_L\rightarrow 3\pi^0$ events.
The similarity of the signal and normalization channels
serves to minimize acceptance and efficiency uncertainties,
and the remaining systematic uncertainties are mainly
external, deriving from the rate of Dalitz decays
and the parametrization of the pion form factor.
The KTeV Collaboration report the value
\bdm
\BRmeas(\decay)
= (7.48 \pm 0.29 \pm 0.25) \times 10^{-8}
\edm
after extrapolating from the selected to the entire
kinematic region. The first error is from data statistics alone, while
the second is the total systematic error.
\par
The most recent theoretical estimate of the $\pion \to \epem$
width was completed by Dorokhov and Ivanov, who obtained
\bdm
\BRSM(\decay)
= (6.2 \pm 0.1) \times 10^{-8} ~,
\edm
and noted the discrepancy with respect to the KTeV measurement~\cite{Dorokhov}.
The dominant theoretical uncertainty comes from the hadronic form factors,
expressed through a dispersion relation in terms of a subtraction constant.
The authors of~\cite{Dorokhov} estimate the value of the subtraction constant
by using an assumed monopole functional form and comparing with CLEO data.
They further compare their estimate with results derived from the operator
product expansion, QCD sum rules, generalized vector meson dominance, and a
non-local constituent quark model.  All results agree with each other within
the quoted uncertainties.
\par
The excess of $\BRmeas$ over $\BRSM$ suggests that
non-SM processes may be contributing to this rare decay.
If the $U$ boson couples to
quarks as well as electrons, the lowest-order contribution to $\decay$
would come from the tree-level process $\pion \to U^{\star} \to \epem$.
The smallness of this contribution would be explained by
very small values of the coupling constants, which are, in fact,
natural in the light dark matter model~\cite{BoehmFayet,FayetLatest}.
\par
The $U$ boson coupling to quarks and electrons can be written in terms of
vector and axial-vector components,
\begin{eqnarray}
{{\cal L}} & \supset &
U_\mu \left\{ \bar{u} \gamma^\mu \left( g_V^u + \gamma_5 g_A^u \right) u
+ \bar{d} \gamma^\mu \left( g_V^d + \gamma_5 g_A^d \right) d \right.
\nonumber \\ & &
\left. + \bar{e} \gamma^\mu \left( g_V^e + \gamma_5 g_A^e \right) e \right\}
\end{eqnarray}
where $u$ and $d$ are the up and down quark fields, and $e$ is the electron field.
It is not necessary to have family-universal couplings, and in fact we will
assume that couplings to the second and third generations are suppressed.
To respect the unitary bound in the ultra-violet, the $U$ should correspond
to a local $U(1)_U$ symmetry, which is spontaneously broken.  One might worry that
the presence of axial vector couplings implies that the Yukawa interactions
between $u$, $d$, and $e$ and the Higgs responsible for generating fermion
masses are not symmetric under $U(1)_U$.  However, given the tiny $u$, $d$,
and $e$ masses compared to the electroweak scale, it is easy to accommodate them
from effective higher dimensional operators induced by high mass states.
\par
\begin{figure}
\hspace*{-4cm}
\vspace*{-1cm}
\includegraphics{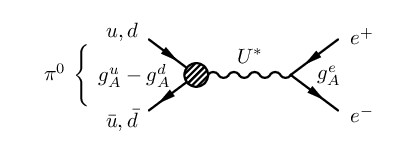}%
\hspace*{-4cm}
\vspace*{0.5cm}
\caption{Feynman diagram for $\decay$. \label{fig:pi0decay}}
\end{figure}

At tree level, the contribution to $\decay$ is mediated by an
off-shell $U$ boson, as depicted in Fig.~\ref{fig:pi0decay}.
The $U$ boson contribution to the matrix element is given by
\begin{eqnarray}
\mathcal{M}_U & = &  \frac{(g_A^d - g_A^u)
g_A^e f_\pi}{m_U^2}[\bar{u}\gamma^\mu \gamma_5 v]p_\mu
\label{eq:Mu}
\end{eqnarray}
where $m_e$, and $m_U$ are the electron, and $U$-boson
masses, $f_\pi$ is the pion decay constant, and $p_\mu$ is the 
$\pi^0$ four-momentum, $p^2 = m_\pi^2$.
(See the Appendix~\ref{S:app} for details).  

To obtain the full amplitude for $\pi^0 \rightarrow e^+ e^-$,
the $U$ boson matrix element is combined with the Standard Model
amplitude for $\pi^0 \rightarrow e^+e ^-$ \cite{Dorokhov} and summed
over the outgoing electron and positron spins.  The partial width
$\pi^0 \rightarrow e^+ e^-$ is computed from
the expression for the two-body decay,
\begin{equation}
\Gamma = \frac{|\vec{p}|}{8\pi m_{\pi}^2}\overline{|\mathcal{M}_{SM} + \mathcal{M}_U|^2}
\label{eq:width}
\end{equation}
where $|\vec{p}|$ is the three-momentum of one of the outgoing particles, and
is equal to approximately $m_\pi / 2$, neglecting the electron mass.

\section{\label{S:quarks}Bounds on \emph{U}-quark couplings}
\par
\begin{figure}
\includegraphics[width=0.5\textwidth]{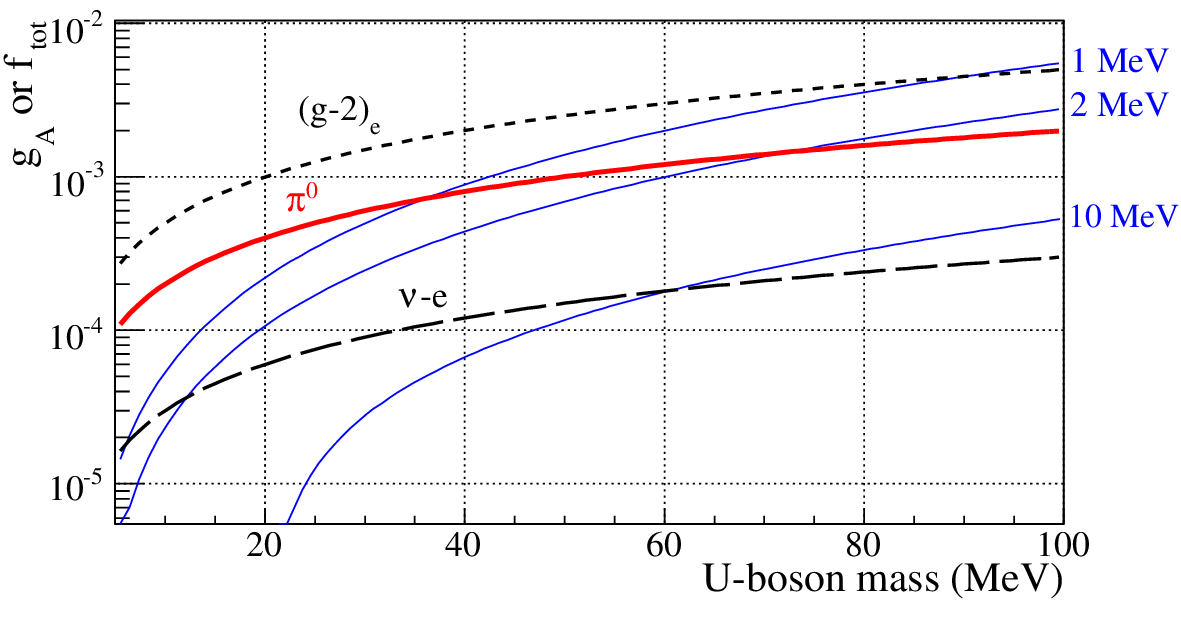}
\caption{\label{F:constraints}
Constraints on the couplings as a function of~$m_U$.
The thick (red) line labeled ``$\pion$'' shows our result Eq.~(\ref{E:constraint}).
The dotted line comes from constraints on $(g-2)_e$ and the dashed line
from $\nu$-$e$ scattering~\cite{FayetLatest}.  The
solid (blue) lines labeled with ``$n$~MeV'' are based
on relic density calculations~\cite{FayetLatest}.
}
\end{figure}

We interpret the positive difference $\BRmeas-\BRSM = (1.3\pm 0.4) \times 10^{-8}$
as the contribution of ${\cal M}_U$ in Eq.~(\ref{eq:width}).  Taking the known pion and electron
masses, $f_{\pi^0} = 130\pm 5$~MeV and $\tau_{\pi^0} = (84\pm 6)\times 10^{-18}$~s~\cite{PDG},
we find
\beq
\frac{(g^u_A - g_A^d) {g^e_A}}{m_U^2} = (4.0 \pm 1.8) \times 10^{-10}~{\rm MeV}^{-2} .
\label{E:constraint}
\eeq
In order to make contact with other constraints on this model, we
assume, as an illustration, that the electron coupling and the
difference in quark couplings are equal, {\em i.e.},
$g^u_A - g_A^d = g^e_A \equiv g_A$.  This choice is arbitrary,
but one might naturally expect such a relation to hold within
an order of magnitude; a more precise relation requires
a specific model for the fermion charges under $U(1)_U$, which
is beyond the scope of this Letter.  With this assumption,
\beq
 g_A = 2.0^{+0.4}_{-0.5} \times 10^{-4} \times \left(\frac{m_U}{10~{\mathrm{MeV}}}\right)
\label{eq:g_A}
\eeq
where the asymmetric error bars come from taking the square root of Eq.~(\ref{E:constraint}).
Fig.~\ref{F:constraints} shows this constraint as a thick line labeled~``$\pion$''.
If a given model specifies a different relation between $g^u_A - g_A^d$ and $g^e_A$,
then this line will move vertically in the plot.
\par
Fayet has derived other bounds on the coupling of $U$ bosons to quarks and leptons
from a variety of processes~\cite{FayetLatest}, and some of these are shown
in Fig.~\ref{F:constraints}.  The dashed line labeled ``$(g-2)_e$''
indicates his constraints on the axial coupling of $U$ to electrons
derived from measurements of the anomalous magnetic
moment of the electron; the region above this line is excluded.
Constraints from kaon decays,
as well as $(g-2)_\mu$~\cite{FayetLatest}, can be evaded if we assume that couplings to second and third generation
fermions are suppressed.  
Neutrino-electron scattering can provide a relatively severe constraint \cite{FayetLatest},
but may be evaded if the coupling to electrons is largely right-handed.
Finally, the three solid lines labeled ``1~MeV,'' etc., show constraints on
the total $U-e$ coupling $f_{tot}~=~\sqrt{(f_V^e)^2 + (f_A^e)^2}$ from
the dark matter relic density~\cite{FayetLatest}, assuming $\Cchi = 1$, for three hypothetical values
of the $\chi$~mass.  The regions above these lines correspond to smaller values of $\Cchi$.
\par
The curves in Fig.~\ref{F:constraints} show that our values for
the couplings of the $U$-boson to light quarks and leptons are
interesting in the context of the light dark matter model, falling
in the same order-of-magnitude as other constraints.
Since $\cal{M}_U$ depends on a set of coupling constants
different from the other constraints, the rare decay~$\decay$
provides a different view of the phenomenology of the light $U$ boson.
\par
We note that the partial width for the electroweak interaction process
$\pion \to Z^{\star} \to \epem$
is identical to Eq.~(\ref{eq:Mu}) with
$m_U$ replaced by $m_Z$ and $g_A^{u,d,e}$
replaced by the analogous weak couplings; even with $g_A^{u,d,e}$ as small~$10^{-4}$,
the large mass of the $Z$ boson renders this electroweak
process completely negligible compared to the $U$-mediated decay.

\section{\label{S:improvements}Possible Improvements}
\par
The excess depends on one precise measurement and one precise calculation,
and has a significance of only $3\sigma$.  Improvements to this situation
would help sharpen the discussion.
\par
The KTeV measurement depends crucially on the branching ratio for Dalitz decays,
which was last measured in 1981~\cite{Schardt} to a precision of~$3\%$.
Combined with a bubble chamber measurement from 1961~\cite{Samios},
the world's best value for this basic benchmark decay is precise
at the level of~$2.7\%$~\cite{PDG}.  We speculate that modern experiments
could provide a more precise measurement.  For example, Na48/2 has
reconstructed $9.1\times 10^7$ decays $K^\pm\rightarrow\pi^\pm2\pi^0$
as a key part of their program to study CP violations in charged kaon
systems~\cite{Na48}.   Their trigger does not appear to have suppressed
Dalitz decays.   As these decays take place in a long vacuum tube,
there is no background from converted photons.  The resolution on photon
energies and position is excellent, allowing a direct reconstruction of
the vertex position.  Combined with the charged pion, the mass resolution
is~$0.9$~MeV$/c^2$.  Track and photon reconstruction efficiencies have
been measured directly from the data.  Perhaps the acceptance ratio for
one charged track plus five electromagnetic clusters over one charged
track plus four electromagnetic clusters could be estimated by simulations
at the $1\%$~level.  If so, a much better measurement of $\BR(\dalitzdecay)$
would appear to be possible.  Clearly, other experiments might endeavor
to update this branching ratio measurement.
\par
Another approach to testing the SM with better precision might be
to compare the measurement of the ratio of branching ratios,
$\BR(\decay)/\BR(\dalitzdecay)$, to the theoretical prediction.
Clearly this ratio is experimentally much more precise, and
we expect the theoretical uncertainties coming from the hadronic
form factor to be reduced as well.
\par
Of course, a direct measurement of $\BR(\decay)$
by another experiment would be extremely interesting.
\par
We considered possible enhancements to leptonic $\eta$ decays
provided by a light $U$-boson.  Following the same calculation
as above, we find
\bdm
{\cal M}_U =  \frac{3 \left(g_A^\ell
(g_A^u + g_A^d - 2 g_A^s) f^8_\eta \right)}{m_U^2}
[\bar{u}\gamma^\mu \gamma_5 v]p_\mu
\edm
Assuming that the couplings are of the same order as indicated
in Eq.~(\ref{eq:g_A}), we find
$\BR(\eta\rightarrow\epem) \sim 10^{-9}$, which is of the same
order as the improved unitary bound~\cite{Dorokhov}, and much smaller
than the experimental bound.  $\eta$~mesons are heavy enough to
decay to muons, and we estimate $\BR(\eta\rightarrow\mpmm) \sim 2.0\times 10^{-5}$.
This prediction is nearly an order of magnitude larger than the measured
value~\cite{Saturn}, $\BR(\eta\rightarrow\mpmm) = (5.7\pm 0.9)\times 10^{-6}$.
If the $U$-boson exists, then its couplings to muons must be smaller
than its couplings to electrons, or the combination of quark axial
couplings $(g_A^u + g_A^d - 2 g_A^s)$ is smaller than $(g_A^u - g_A^d)$.

\section{\label{S:conc}Summary}
\par
The $3\sigma$ excess of the KTeV measurement of $\BR(\decay)$
over the most recent calculation by Dorokhov and Ivanov prompts
considerations of a new physics contribution to this SM-suppressed
decay.  The light neutral vector $U$ boson proposed in the
light dark matter model provides a good basis for calculating
such a contribution.  We carried out such a calculation and
find that couplings of the $U$-boson to electrons and light
quarks should be on the order of $2\times 10^{-4}$,
for $m_U = 10$~MeV.  Such small couplings are consistent with
the expectations of the light dark matter model, and with
other constraints coming from leptonic processes.

\appendix*
\section{\label{S:app}Matrix element calculation}

The $U$-boson exchange couples to both vector and axial vector currents of
quarks, evaluated between a $\pi^0$ state and the vacuum
(see Figure~\ref{fig:pi0decay}),
\bdm
\langle 0 | \left\{
 \bar{u} \gamma^\mu \left( g_V^u + \gamma_5 g_A^u \right) u
+ \bar{d} \gamma^\mu \left( g_V^d + \gamma_5 g_A^d \right) d
\right\} | \pi^0 \rangle ~.
\edm
We can reorganize the four terms into combinations with definite parity and
transformation under strong iso-spin, $SU(2)_V$.  From there, it is easy to
see that the only combination which contributes is the current proportional to
the third component of axial iso-spin,
\bdm
\langle 0 | \frac{1}{2} \left\{
 \bar{u} \gamma^\mu \gamma_5 u - \bar{d} \gamma^\mu \gamma_5 d
\right\} | \pi^0 \rangle \equiv f_\pi p^\mu~.
\edm
where $p^\mu = p_{e^+}^\mu + p_{e^-}^\mu$ is the momentum of the pion.  Note
that the pion state
singles out the axial couplings in the combination $g_A^u - g_A^d$.

The hadronic vertex is contracted through the $U$ boson, which also carries
momentum $p$ into the leptonic vertex,
\[
\mathcal{M} = \frac{(g_A^u - g_A^d) g_A^e}{m_{\pi}^2 - m_U^2}
[\bar{u}\gamma^\mu \gamma_5 v]
\left [g_{\mu \nu} - \frac{p_{\mu} p_{\nu}}{m_U^2} \right ]f_{\pi}p^\nu
\]
where $u$ and $v$ are the electron and positron spinors, respectively.
For an on-shell pion,
$p^\mu p_\mu = m_{\pi}^2$, and we obtain
\[
\mathcal{M} = \frac{(g_A^d - g_A^u)
g_A^e f_\pi}{m_U^2}[\bar{u}\gamma^\mu \gamma_5 v]p_\mu ~.
\]

\begin{acknowledgments}
\par
We thank C\'eline Boehm and for stimulating discussions and
Mayda Velasco for interesting discussions of measurements of light meson decays.

Research at Argonne National Laboratory is supported in part by the
Department of Energy under contract DE-AC02-06CH11357, and the work
by Y.K. and M.S. is supported by contract DE-FG02-91ER40684.

\end{acknowledgments}

\end{document}